# The formation of circumbinary planets through disc fragmentation

Matthew Teasdale ★ and Dimitris Stamatellos ★

*Jeremiah Horrocks Institute for Mathematics, Physics and Astronomy, University of Lancashire, Preston PR1 2HE, UK*

## ABSTRACT

Over 50 circumbinary exoplanets have been discovered in recent years, with several of them being gas giants on wide orbits (>10 au). The aim of this work is to investigate whether these planets can form through circumbinary disc fragmentation due to gravitational instability. We perform hydrodynamic simulations of marginally unstable (i) circumstellar discs, (ii) circumbinary discs with the same temperature profile as the circumstellar discs (fiducial model), and (iii) realistic circumbinary discs heated individually by each star of the binary. We find that discs around binaries with wider separations fragment earlier and more efficiently than those around closer binaries, and earlier than circumstellar discs. Realistic circumbinary discs form a larger number of protoplanets ($9 \pm 0.9$ protoplanets per disc), than fiducial circumbinary ($6.5 \pm 0.6$), and circumstellar discs ($7.5 \pm 0.8$). In realistic circumbinary discs, initial protoplanet masses are lower than those formed in circumstellar discs, and a larger fraction of them lie in the planetary-mass regime, favouring the formation of gas giant planets over brown dwarfs or low-mass stars. Fragmentation occurs predominantly beyond a binary-imposed forbidden region of ~50 au, leading to final orbital radii peaking at ~100 au. We also find that in circumbinary discs dynamical interactions eject a higher fraction of protoplanets than in circumstellar discs, producing free-floating objects, with ejection velocities on the order of 2-6 km s$^{-1}$. We conclude that gravitational fragmentation of circumbinary discs is a viable and potentially significant formation pathway for circumbinary gas giant planets.

## 1 INTRODUCTION

It is now well established that approximately half of all stars are in binary or higher order multiple systems, with up to 30–40 per cent of these (dependent on the stellar mass) being close systems with separations of less than 10 au (G (D. Raghavan et al. 2010); Duchêne et al. 2013; M. Moe & R. Di Stefano 2017; S. S. R. Offner et al. 2023). It is expected that close binaries are attended by circumbinary discs, like e.g. GG Tau (S. Guilloteau, A. Dutrey & M. Simon 1999), HD 142 527 (M. Fukagawa et al. 2006; A. P. Verhoeff et al. 2011), L1448 IRS3B (J. J. Tobin et al. 2016; N. K. Reynolds et al. 2021), VLA 1623A (R. J. Harris et al. 2018; S. I. Sadavoy et al. 2024; I. C. Radley et al. 2025), and L1551 IRS 5 (N. Cuello et al. 2026).

Since the discovery of the exoplanet Kepler-16b (L. R. Doyle et al. 2011), over 50 circumbinary exoplanets have been confirmed (J. L. Christiansen et al. 2025); these have presumably formed in circumbinary protostellar discs (B. Quarles et al. 2018; A. B. T. Penzlin et al. 2024). Circumbinary planets are diverse in semimajor axis, mass and eccentricity (see Fig. 1). They have masses ranging from 0.006 to 26 Jupiter masses and distances from their host binary from 0.2 to 1100 au. However, the majority orbit less than 10 au from the binary. This may be due to obser-vational biases and it is likely that more wide-orbit circumbinary planets are yet to be discovered.

Giant planets are thought to form either by core accretion or by disc fragmentation due to gravitational instability. The core accretion model states that a core forms through pebble and/or planetesimal accretion within a gaseous disc (P. Goldreich & W. R. Ward 1973; H. Mizuno 1980; P. Bodenheimer & J. B. Pollack 1986; J. B. Pollack et al. 1996; J. Drążkowska et al. 2023). If the core gains a critical mass of around ~10 M$_\oplus$, a gaseous envelope may be attained. This process has difficulty forming gas giant planets on wide orbits due to the growth timescale, ~10 Myr (J. B. Pollack et al. 1996), being at odds with the dispersal of the protostellar disc (~3-5 Myr; K. Wagner, D. Apai & K. M. Kratter 2019). The formation of gas giant planets is also possible through the gravitational instability model. A disc may become gravitationally unstable if it is massive enough and/or cool enough so that the Toomre criterion is satisfied (A. Toomre 1964),

$$Q = \frac{c_s \Omega}{\pi G \Sigma} \lesssim Q_{crit} \simeq 1 - 2, \qquad (1)$$

where $Q$ is the Toomre parameter, $c_s$ is the sound speed, $\Omega$ the angular frequency, $G$ the gravitational constant, and $\Sigma$ the surface density of the disc. A potential outcome of gravitational instability is fragmentation. This may lead to the formation of gas giant planets, should the cooling time of the disc be sufficiently short, i.e. $\tau_c \lesssim 3\Omega^{-1}$ (C. F. Gammie 2001; B. M. Johnson & C. F. Gammie 2003; W. K. M. Rice et al. 2003; W. K. M. Rice, G. Lodato & P. J. Armitage 2005; F. Meru & M. R. Bate 2012; H. Deng, L. Mayer

* E-mail: MTeasdale1@lancashire.ac.uk (MT); DStamatellos@lancashire.ac.uk (DS)

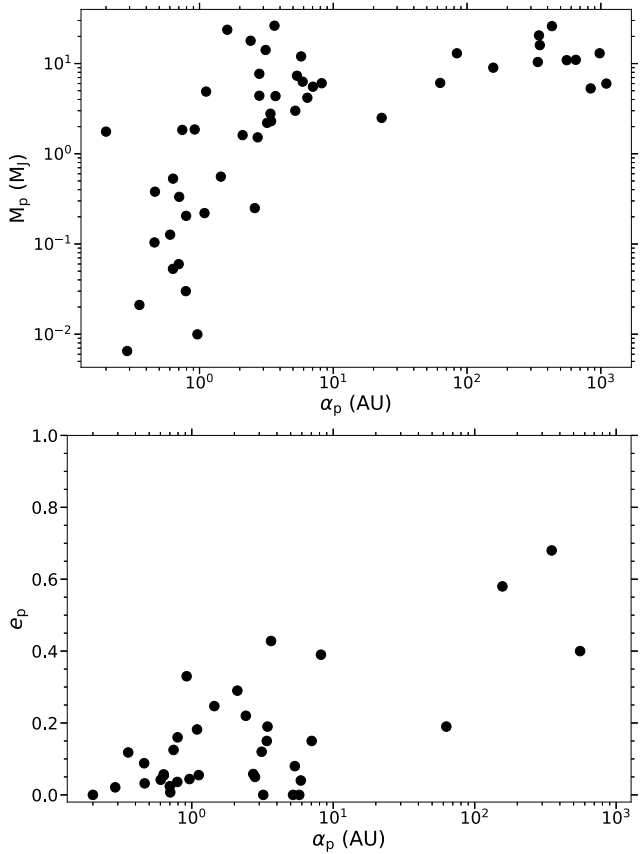

**Figure 1.** The separation of the currently known circumbinary planets from their host binaries, plotted against their mass (top; a) and eccentricity (bottom; b). Data from of The Extrasolar Planets Encyclopaedia (https://exoplanet.eu) and from the NASA Exoplanet Archive (https://exoplanetarchive.ipac.caltech.edu) which is operated by the California In-stitute of Technology, under contract with the National Aeronautics and SpaceAdministration under the Exoplanet Exploration Program (J. L. Christiansen et al. 2025).

& F. Meru 2017). The conditions listed above may be satisfied at large disc radii where fragmentation is likely (A. C. Boley 2009; D. Stamatellos & A. P. Whitworth 2009a).

Circumbinary planets form the same way as the circumstellar ones, with the complication that the presence of the binary and its non-axisymmetric gravitational potential creates a dynamically unstable, forbidden region where planet cannot exist (M. J. Holman & P. A. Wiegert 1999; F. Marzari & P. Thebault 2019; A. Langford & L. M. Weiss 2023). Many of the observed circumbinary exoplanets are observed near the edge this forbidden region, although an *in situ* formation is thought to be unlikely (F. Marzari & H. Scholl 2000; F. Marzari, P. Thébault & H. Scholl 2008; S.-J. Paardekooper & Z. M. Leinhardt 2010; E. L. Martín et al. 2013); these planets probably formed farther away and migrated towards the the edge of the circumbinary cavity, which is expected to be 2–3 times the binary separation (P. Artymowicz & S. H. Lubow 1994; A. Pierens & R. P. Nelson 2008; W. Kley & N. Haghighipour 2015; W. Kley, D. Thun & A. B. T. Penzlin 2019). G. A. L. Coleman, R. P. Nelson & A. H. M. J. Triaud (2023) and G. A. L. Coleman (2024) demonstrate that circumbinary planets may form by pebble accretion farther out in the disc and subsequently migrate to the locations observed. On the other hand the formation of cir-cumbinary planets by fragmentation of gravitationally unstable discs has not been studied so far.

Recently, M. Teasdale & D. Stamatellos (2026) investigated the lower mass limit for circumbinary disc fragmentation. They find that realistic circumbinary discs fragment at lower masses than circumstellar discs, Specifically, they find that a disc around a binary with a combined mass of $0.7\,M_{\odot}$ fragments down to a disc-to-stellar mass ratio of 0.17-0.26, compared to a ratio of 0.31 needed for a disc around a single star with the same mass. This corresponds to a minimum disc mass that is 45 per cent lower, suggesting that planet formation by disc fragmentation may be easier around circumbinary discs.

Here, we extend the work of M. Teasdale & D. Stamatellos (2026). The aim of this work is to study the statistical properties of protoplanets formed through gravitational instability and subsequent fragmentation of circumbinary discs, and to compare them with those formed in circumstellar discs. We use the term *protoplanets* to refer to all objects that form in the disc, irrespective of their mass; some of them will end up as planets but others they may grow further in mass to become brown dwarfs.

Previous studies of fragmentation of circumstellar discs (D. Stamatellos, D. A. Hubber & A. P. Whitworth 2007a; D. Stamatellos & A. P. Whitworth 2009a; A. Mercer & D. Stamatellos 2017, 2020; A. Fenton & D. Stamatellos 2024) have shown that disc fragmentation leads to the formation of wide-orbit planets (but also to the formation of brown dwarfs and low-mass stars). Initially, most objects that form have masses of a few times that of Jupiter (as expected due to the opacity limit for gas fragmentation; M. J. Rees 1976; A. P. Whitworth & D. Stamatellos 2006). However, these protoplanets can rapidly accrete gas, growing in mass to become brown dwarfs or low-mass hydrogen-burning stars (D. Stamatellos & A. P. Whitworth 2009a; K. M. Kratter et al. 2010a; E. I. Vorobyov 2013; K. Kratter & G. Lodato 2016; D. Stamatellos & S.-i. Inutsuka 2018), although it is also possible that low-mass planets form (S. Nayakshin 2017; H. Deng, L. Mayer & R. Helled 2021). A fraction of the objects formed by fragmentation may be ejected from the disc due to dynamical interactions, becoming free-floating (J. N. Winn & D. C. Fabrycky 2015). We investigate when and where planets form in circumbinary discs, what their masses and orbital properties are, and what fraction of these planets are ejected into the field.

The paper is structured as follows. In Section 2, we describe the computational method and simulation set-up. In Section 3, we present the properties of the protoplanets formed by fragmentation of circumbinary discs. Finally, we discuss the wider implications of this work and present our conclusions in Section 4.

## 2 COMPUTATIONAL METHOD AND SIMULATION SET-UP

We use the computational method described in M. Teasdale & D. Stamatellos (2026) to investigate the properties of protoplanets formed through gravitational instability in circumbinary discs. We simulate gaseous circumstellar, and circumbinary discs using SEREN, a 3D SPH code developed by D. A. Hubber et al. (2011). The code employs the polytropic radiative transfer approximation (D. Stamatellos et al. 2007b; J. C. Lombardi, W. G. McInally & J. A. Faber 2015; A. K. Young et al. 2024).

We perform three *sets* of simulations (see M. Teasdale & D. Stamatellos 2026, for details). The first set of simulations covers circumstellar discs heated by a single star (referred to as the *circumstellar* model; CS), the second set covers circumbinary discs with the same temperature profile as the circumstellar disc model (referred to as the *fiducial* model; CBF), and the third set covers circumbinary discs that are heated by each star individually (referred to as the *realistic* model; CBR). The binary parameters are seen in table 1 of M. Teasdale & D. Stamatellos (2026). We model binaries with separations 5 and 10 au, and with eccentricities 0.2 and 0.5. For each combination of binary parameters we perform a number of simulations with different realizations of the disc, to account for the stochastic nature of disc fragmentation. In total we perform 13 CS simulations, 21 CBF, simulations and 12 CBR simulations (the numbers were chosen so as to have approximately 100 objects formed per set of simulations).

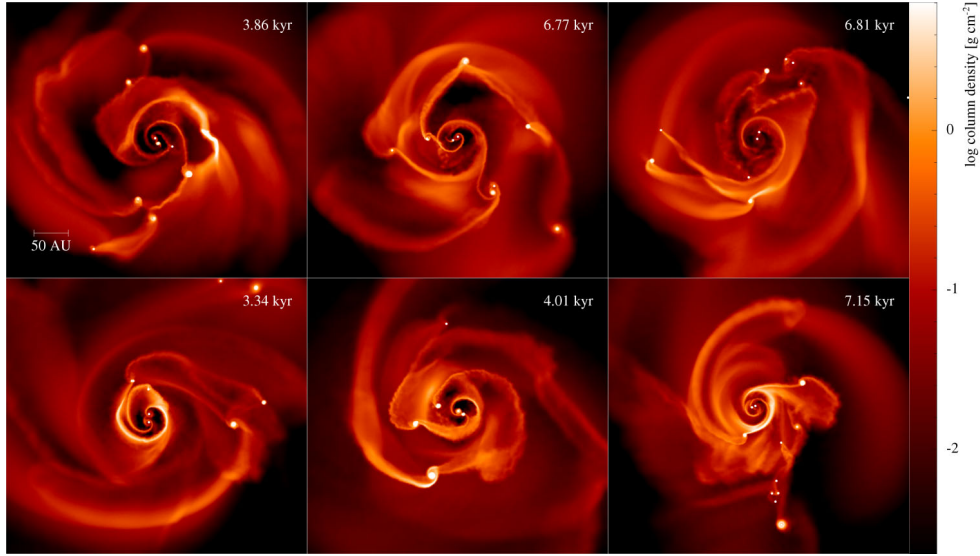

**Figure 2.** Disc surface density (g cm$^{-2}$) snapshots for representative circumbinary fiducial model simulations with a disc mass of $M_D = 0.22\,M_\odot$, and a binary separation of $\alpha_b = 10$ au at the end of the simulation run time (i.e. when 70 per cent of the disc mass has been accreted). The first column corresponds to simulations with a binary mass ratio of $q_b = 1$, the second with a binary mass ratio of $q_b = 0.3$, and the third with a binary mass ratio of $q_b = 0.1$. The top row corresponds to simulations with a binary eccentricity of $e_b = 0.2$ and the bottom row with a binary eccentricity of $e_b = 0.5$.

We assume a disc, represented by $10^6$ SPH particles, extending from $R_D^{in} = 5$ au to $R_D = 120$ au. We use a stellar mass of $M_\star = 0.7\,M_\odot$ for the circumstellar case, which is also the total mass of the binary in the two other sets of simulations. In M. Teasdale & D. Stamatellos (2026) we find that the minimum disc mass for fragmentation is close to $0.22\,M_\odot$ for the CS and CBF models, and $0.18\,M_\odot$ for the CBR model (depending on the binary properties).

## 3 FRAGMENTATION OF CIRCUMBINARY DISCS

The discs in all three sets of simulations (CS, CBF, and CBR) become gravitationally unstable and fragment (see Figs 2–5 for snapshots from representative simulations). We assume that a protoplanet has formed when a condensation in a disc reaches a density of $10^{-9}$ g cm$^{-3}$, whereby a sink with an accretion radius of $R_s = 0.1$ au is introduced to represent the protoplanet. We allow the disc to evolve until 70 per cent of the material has been accreted onto either the star(s) or the protoplanets (D. Stamatellos & A. P. Whitworth 2009a).

Some of the protoplanets remain bound to the central star(s), whereas others are ejected due to planet–planet and/or planet–binary interactions. Table 1 summarizes the number of proto-planets formed for each parameter combination and the number of the ones ejected.

### 3.1 The evolution of binary properties

Before discussing the protoplanet properties, we briefly examine how the properties of the binaries change as the discs evolve.

Fig. 6 shows the properties of the binaries at the end of the simulations. We see little to no change in the binary mass ratios for all simulations (see Fig. 6a), with some binaries increasing their mass ratio, due to the secondary accreting more material from the inner disc edge of the disc than the primary.

For binaries with an initial separation of 5 au, most systems become tighter, with final separations in the range 4.5–5.5 au. Binaries with an initial separation of 10 au exhibit both increases and decreases in separation, resulting in final separations between 8–11 au for most of the systems.

In general, we find that most binaries evolve away from their initial eccentricities (see Fig. 6b), settling to values in the range 0.2–0.5. Consequently, binaries with an initial eccentricity of $e_b = 0.2$ typically experience an increase in eccentricity, whereas those with an initial eccentricity of $e_b = 0.5$ generally show a decrease.

These results are consistent with previous studies of binaries attended by circumbinary discs that show that the orbital evolution of the binary depends on the specifics of the system, like e.g. disc thickness, viscosity, and self-gravity (P. Artymowicz & S. H. Lubow 1994; R. Miranda, D. J. Muñoz & D. Lai 2017; D. J. Muñoz, R. Miranda & D. Lai 2019; R. M. Heath & C. J. Nixon 2020; C. Chen, P. J. Armitage & C. J. Nixon 2026).

### 3.2 Protoplanet properties

In total, 341 protoplanets form across all three disc models (see Table 1). The CS model forms 97 protoplanets, the CBF model forms 136 protoplanets and the CBR model forms 108 protoplanets. Fig. 7 shows the average number of protoplanets per disc formed for each model, and for each individual combination of binary parameters. The CBR model is the most efficient one in forming protoplanets Fig. 7a), despite being the one with the lower disc mass (we note that discs in all three models are marginally unstable; see Section 2). This is due to the disc being cooler, which allows the binary to exert a more pronounced influence on the development of the gravitational instability. More protoplanets formed per disc around wider binaries, but there seems to be no dependence on the number of protoplanets on the binary mass ratio or eccentricity (Figs 7b, c, d).

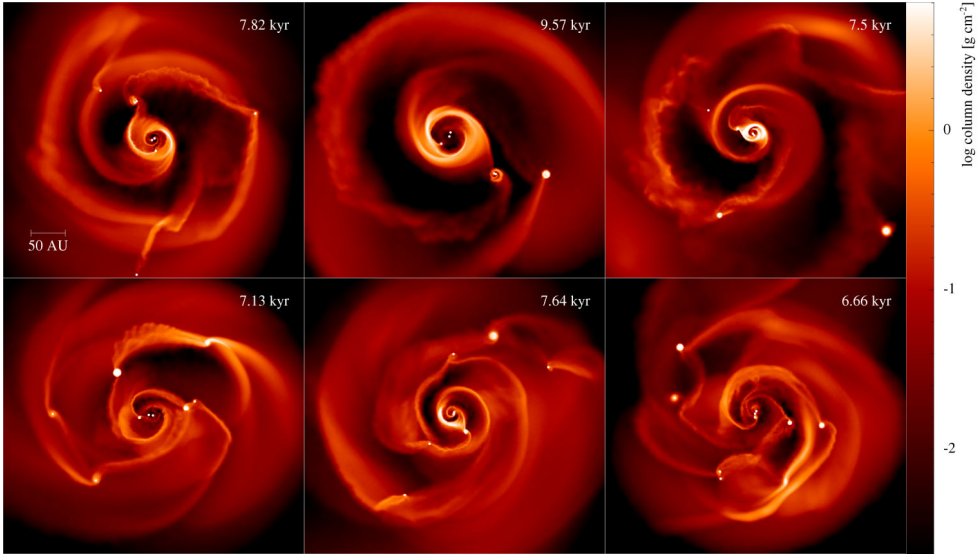

**Figure 3.** Same as in Fig. 2 but for CBF simulations with an initially binary separation of $\alpha_b = 5$ au.

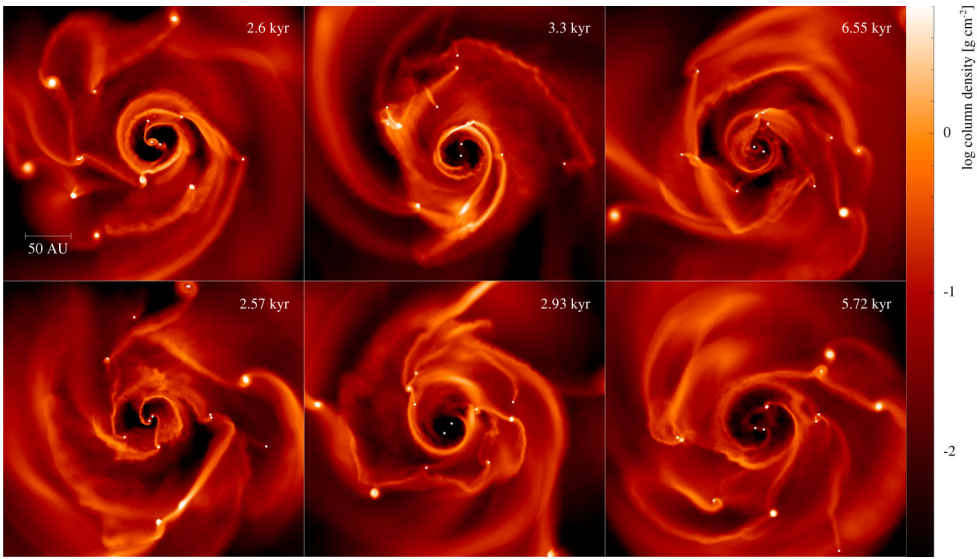

**Figure 4.** Disc surface density (g cm⁻²) snapshots for representative circumbinary realistic model simulations with a disc mass of $M_D = 0.22\,M_\odot$, and a binary separation of $\alpha_b = 10$ au at the end of the simulation run time (i.e. when 70 per cent of the disc mass has been accreted). The first column corresponds to simulations with a binary mass ratio of $q_b = 1$, the second with a binary mass ratio of $q_b = 0.3$, and the third with a binary mass ratio of $q_b = 0.1$. The top row corresponds to simulations with a binary eccentricity of $e_b = 0.2$ and the bottom row with a binary eccentricity of $e_b = 0.5$.

### 3.2.1 Mass, orbital radius, and formation time

Fig. 8 shows the formation mass, orbital radius and time of the protoplanets formed across all three models. Protoplanets form over a wide range of orbital radii ($>20$ au) and they initially have typical masses from 1 to $4\,M_J$ (see Fig. 8a), which is as expected from the opacity limit for gas fragmentation (M. J. Rees 1976; A. P. Whitworth & D. Stamatellos 2006; D. Stamatellos & A. P. Whitworth 2009b). However, a few protoplanets forming in the CS model have higher masses (5-14 $M_J$). We see a distinct separation in the formation time of the protoplanets (Fig. 8b, c). Discs around wider binaries ($\alpha_b = 10$ au) and mass ratio $q_b = 1$ and 0.3 fragment faster than discs around binaries with a lower separation and mass ratio (Figs 8b, c), demonstrating the more pronounced influence of the wide binary in disc evolution. This result is further supported by the fact that circumstellar discs tend to fragment at later times than circumbinary discs. There is no correlation between the formation mass and the formation orbital radius nor the formation time (Fig. 8a, b). On the other hand, there is a small tendency for faster formation time (i.e. fragmentation) closer to the star(s) (Fig. 8c).

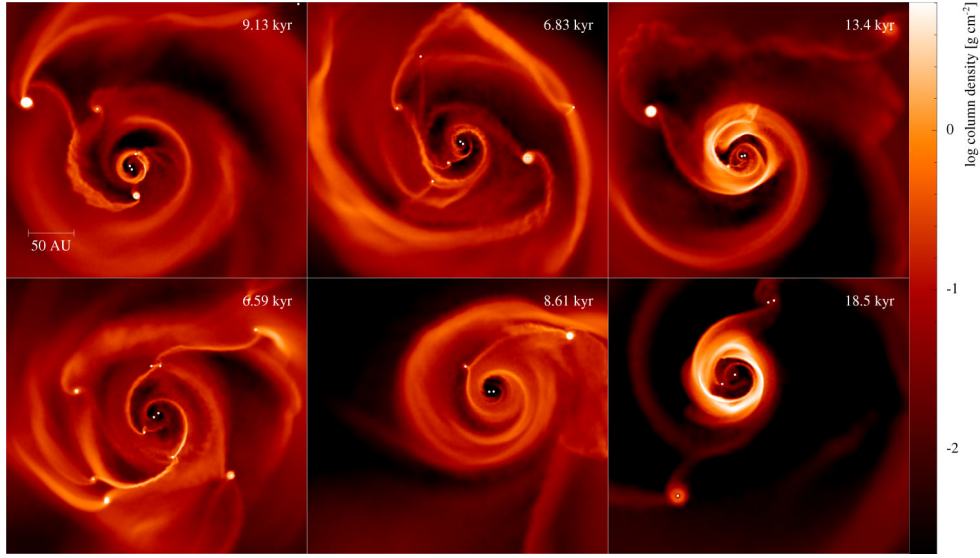

**Figure 5.** Same as in Fig. 4 but for CBR simulations with an initially binary separation of $\alpha_b = 5$ au.

**Table 1.** The number of protoplanets formed for circumstellar, and the circumbinary fiducial and realistic models. Here, Type refers the CS, CBF, CBR models, $\alpha_b$ is the initial binary separation, $q_b$ the binary mass ratio, $e_b$ the binary eccentricity, $R$ is the number of runs, $N_p$ is the number of formed protoplanets, $<N_p>$ is the average number of protoplanets per disc, $E_p$ is the number of ejected protoplanets and $<E_p>$ is the average number of ejected protoplanets per disc.

| Type | $\alpha_b$ (au) | $q_b$ | $e_b$ | $R$ | $N_p$ | $<N_p>$ | $E_p$ | $<E_p>$ |
|---|---|---|---|---|---|---|---|---|
| CS | – | – | – | 13 | 97 | 7.5 | 3 | 0.2 |
| CBF | 10 | 1 | 0.2 | 3 | 19 | 6.3 | 2 | 0.7 |
| | | | 0.5 | 3 | 16 | 5.3 | 1 | 0.3 |
| | | 0.3 | 0.2 | 2 | 14 | 7 | 0 | 0 |
| | | | 0.5 | 3 | 20 | 6.7 | 0 | 0 |
| | | 0.1 | 0.2 | 1 | 9 | 9 | 0 | 0 |
| | | | 0.5 | 1 | 9 | 9 | 0 | 0 |
| | 5 | 1 | 0.2 | 1 | 6 | 6 | 0 | 0 |
| | | | 0.5 | 2 | 13 | 6.5 | 1 | 0.5 |
| | | 0.3 | 0.2 | 1 | 4 | 4 | 0 | 0 |
| | | | 0.5 | 1 | 7 | 7 | 0 | 0 |
| | | 0.1 | 0.2 | 1 | 5 | 5 | 0 | 0 |
| | | | 0.5 | 2 | 14 | 7 | 1 | 0.5 |
| CBF$_{total}$ | – | – | – | 21 | 136 | 6.5 | 5 | 0.2 |
| CBR | 10 | 1 | 0.2 | 1 | 11 | 11 | 0 | 0 |
| | | | 0.5 | 1 | 9 | 9 | 0 | 0 |
| | | 0.3 | 0.2 | 1 | 12 | 12 | 0 | 0 |
| | | | 0.5 | 1 | 12 | 12 | 1 | 1 |
| | | 0.1 | 0.2 | 1 | 13 | 13 | 0 | 0 |
| | | | 0.5 | 1 | 12 | 12 | 0 | 0 |
| | 5 | 1 | 0.2 | 1 | 7 | 7 | 1 | 1 |
| | | | 0.5 | 1 | 11 | 11 | 1 | 1 |
| | | 0.3 | 0.2 | 1 | 7 | 7 | 0 | 0 |
| | | | 0.5 | 1 | 7 | 7 | 2 | 2 |
| | | 0.1 | 0.2 | 1 | 3 | 3 | 0 | 0 |
| | | | 0.5 | 1 | 4 | 4 | 0 | 0 |
| CBR$_{total}$ | – | – | – | 12 | 108 | 9 | 5 | 0.4 |

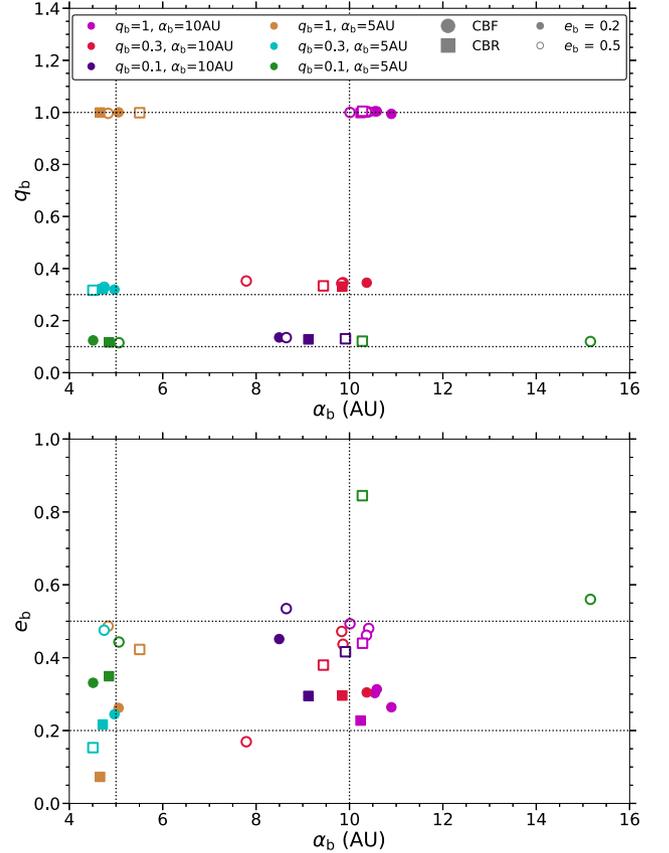

**Figure 6.** The binary separation at the end of the simulations (i.e. when 70 per cent of the disc has been accreted), plotted against the binary mass ratio (top; a) and eccentricity (bottom; b). The dotted lines indicate the initial values of these properties.

Fig. 9 shows the final mass of the protoplanets versus their final orbital radius, formation time, and formation radius. The vast majority of protoplanets remain within their parent discs by the end of the simulation, i.e. within a final radius below ~200 au (Fig. 9a). We also find that protoplanets closer to the star(s) gener-ally attain a larger final mass, as there is a larger amount of gas to accrete from the disc in this region (Fig. 9a). There seems to be no correlation between the formation time and the final protoplanet mass (Fig. 9b). High-mass protoplanets (i.e. above ∼20 $M_J$, BD-mass objects) are more likely to have formed at smaller radii in the disc (Fig. 9c).

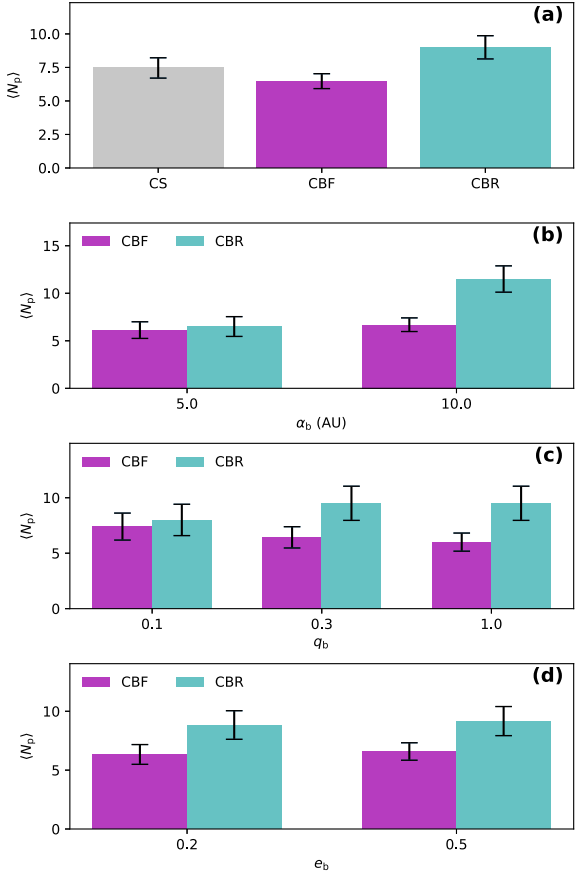

**Figure 7.** (a) The average number of protoplanets formed per disc for each model and its dependence on the binary separation (b), mass ratio (c), and eccentricity (d). See Table 1 for a breakdown of the individual values for each combination of binary parameters.

### 3.2.2 Accretion rates on to protoplanets

Protoplanets are actively accreting at the end of the simulations. Fig. 10, presents the final accretion rate on to the protoplanets versus their mass (a) and orbital radius (b). The accretion rates vary over several orders of magnitude, spanning $10^{-5}$–$20\,M_J\mathrm{yr}^{-1}$, with lower accretions rate seen for lower mass protoplanets (Fig. 10a) and wider orbits (Fig. 10b). These accretion rates are larger than those inferred for observed systems; this may be due to the older ages of the observed systems compared to our simulated protoplanets.

The accretion rates we find are two orders higher than the accretion rates estimated for the embedded planets PDS 70 b,c (M. Keppler et al. 2018; S. Y. Haffert et al. 2019) that orbit a single star; these accrete at ∼$10^{-8}$–$10^{-7}\,M_J\mathrm{yr}^{-1}$ (K. Wagner et al. 2018; J. J. Wang et al. 2020; Y. Zhou et al. 2021). AB Aurigae b (T. Currie et al. 2022, 2025) (again a disc-embedded planet around a single star) accretes at much higher rate that the PDS 70 planets,

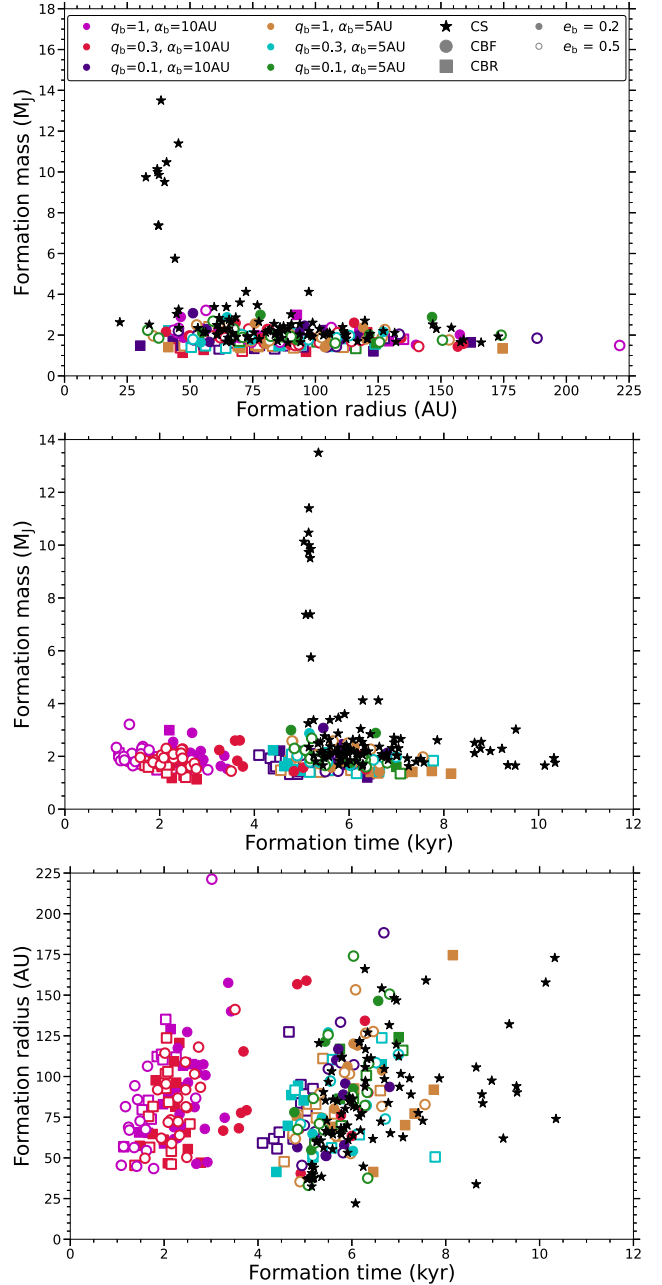

**Figure 8.** The formation mass, radius and time of the protoplanets formed in the circumstellar, circumbinary fiducial, and realistic model discs. The top panel shows the formation orbital radius plotted against the formation mass (a), the middle panel shows the formation time is plotted against the formation mass (b), and the bottom panel shows the formation time is plotted against the formation radius (c).

∼$10^{-6}\,M_J\mathrm{yr}^{-1}$ (T. Currie et al. 2022; Y. Zhou et al. 2022), which is more consistent with our results. WISPIT 2b (J. Li et al. 2025; L. M. Close et al. 2025; R. F. Capelleveen et al. 2025), a wide orbit (∼60 au), gap-embedded planet around a single star, exhibits a much lower accretion rate (∼$10^{-12}\,M_J\mathrm{yr}^{-1}$) (J. Li et al. 2025).

These systems are young, with ages 1-5 Myr. Delorme 1 (AB)b, which is a much older circumbinary planet (∼40 Myr old; P. Delorme et al. 2013), has an estimated accretion rate from 3.4×$10^{-10}\,M_J\mathrm{yr}^{-1}$ to 2.0×$10^{-8}\,M_J\mathrm{yr}^{-1}$ (S. C. Eriksson et al. 2020;

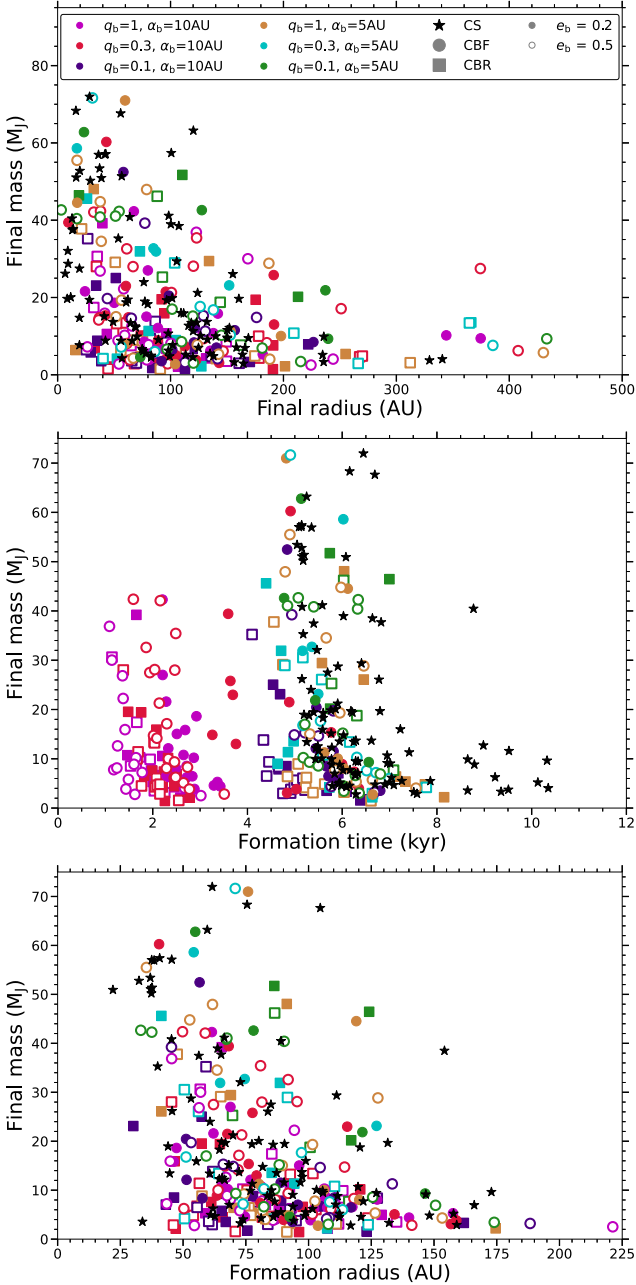

**Figure 9.** The final orbital radius, formation radius and formation time of the formed in the CS, CBF, and CBR models plotted against their final mass when their parent discs have had 70 per cent of the initial mass accreted. The top panel (a) shows the final radius, the middle panel (b) shows the formation time and the bottom panel (c) shows the formation radius of the protoplanets. Several planets are ejected from the system, reaching a final radius ≥1000 au.

S. K. Betti et al. 2022; S. C. Ringqvist et al. 2023; G.-D. Marleau et al. 2024).

We can calculate the protoplanet accretion luminosity corresponding to the estimated accretion rates using

$$L_{\rm p} = f \frac{G M_{\rm p} \dot{M}_{\rm p}}{R_{\rm p}}, \tag{2}$$

where $f$ is the radiative efficiency factor. We assume that, at this early stage, accretion happens onto the second hydrostatic core (D. Stamatellos & A. P. Whitworth 2009b; A. Mercer & D. Stamatellos 2020); therefore we set the accretion radius equal to the size of the second core, i.e. $R_{\rm p} \approx 3\,{\rm R_{\odot}}$. This loosely corresponds to a hot-start planet formation scenario (M. S. Marley et al. 2007). Under these assumptions we obtain typical protoplanet luminosities $10^{-5} - 1\,{\rm L_{\odot}}$ (for a $10{\rm M_J}$ protoplanet and $f = 0.1$). These are relatively high luminosities and in principle can be observed; however, they are short-lived (they last only for a few thousand years) and are expected to considerably reduce over time as the accretion rate drops and the second core contracts quasi-statically.

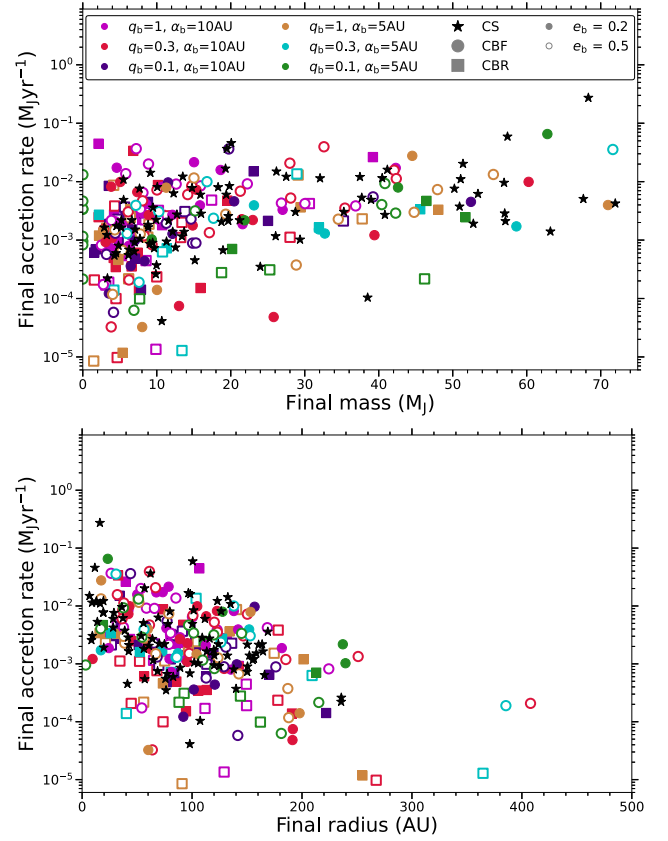

**Figure 10.** The final accretion rate onto the protoplanets plotted against the final masses (top; a) and final orbital radii (bottom; b) in the CS, CBF, and CBR models.

### 3.2.3 Protoplanet migration

Fig. 11 shows the fractional change in the protoplanet's radius between the time the protoplanet forms and at the end of the simulation (when 70 per cent of the initial disc mass has been accreted), i.e.

$$\frac{\Delta r_{\rm p}}{r_{\rm p}} = \frac{r_{\rm p,f} - r_{\rm p,i}}{r_{\rm p,i}}, \tag{3}$$

plotted against the fractional change of the protoplanet's mass during the same period,

$$\frac{\Delta M_{\rm p}}{M_{\rm p}} = \frac{M_{\rm p,f} - M_{\rm p,i}}{M_{\rm p,i}}. \tag{4}$$

Fig. 11(a) presents the inward migrating protoplanets (i.e. when the fractional change in the radius is negative) and Fig. the 11(b) shows the outward migrating ones (i.e. when the fractional change in the radius is positive). 44 per cent of the protoplanets migrate inwards, with 56 per cent moving outwards. Inward migration can bring protoplanets very close to the binary, with most exhibiting a fractional change in radius between 0.1 and 1 (Fig. 11a). On the other hand, outward migration can carry protoplanets to distances up to ∼30 times their formation radius, with most showing fractional changes in radius between 0.1 and 10 (Fig. 11b).

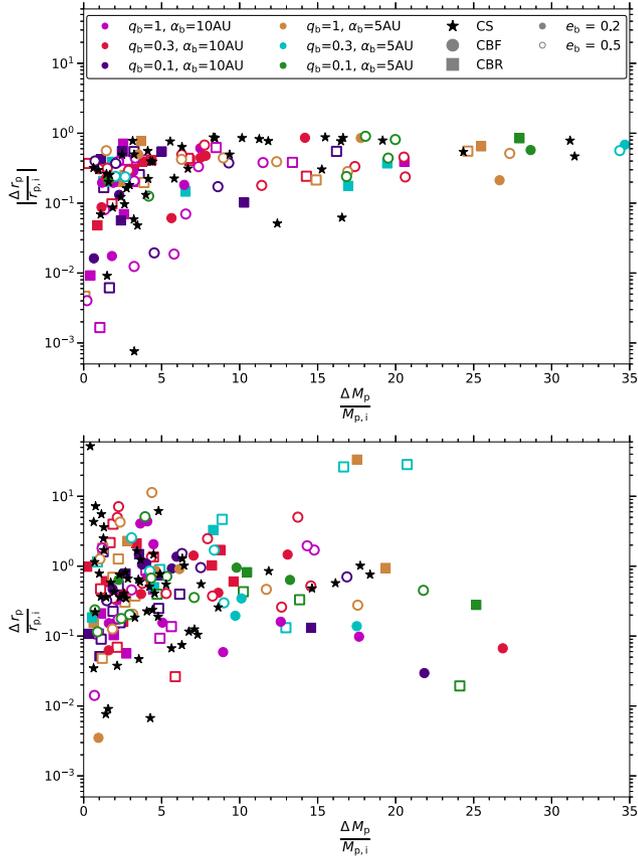

**Figure 11.** The fractional change in the mass of the protoplanets plotted against the fractional change in the radius of the protoplanets, between the time of formation and the end of the simulation. The top panel (a) shows inward migrating protoplanets (i.e. a negative fractional change in the radius of the protoplanets) and the bottom panel (b) shows outward migrating protoplanets (i.e. a positive fractional change in the radius of the protoplanets).

This fraction of inward and outward migrating protoplanets is similar for the CS and CB models, indicating little difference due to the presence of a binary compared to a single star. For the two CB models, the CBF protoplanets are slightly more likely to migrate inwards (46 per cent for the CBF model compared to 42 per cent for the CBR model). Closer binaries ($a_b$ = 5 au) have a larger percentage (61 per cent) of protoplanets migrating outwards, than wider binaries ($a_b$ = 10 au) protoplanets (53 per cent). As the binary mass ratio decreases the fraction of protoplanets migrating inwards drops whereas the fraction of outward migrating protoplanets increases (49 per cent for $q_b$ = 1 compared with 64 per cent for $q_b$ = 0.1). Therefore, a protoplanet forming around a wider separation and higher mass ratio binary is more likely to migrate inwards as opposed to a protoplanet forming around a tighter, smaller mass ratio binary. Interestingly we see no dependence on binary eccentricity, with both modelled showing a larger fraction of protoplanets migrating outwards (56 per cent) compared to inwards (44 per cent).

We find no correlation between changes in mass and changes in orbital radius, indicating that protoplanet migration is not primarily driven by standard disc–planet interactions (A. Pierens & R. P. Nelson 2008; W. Kley & N. Haghighipour 2015; D. Thun & W. Kley 2018; W. Kley et al. 2019; M. Teasdale & D. Stamatellos 2023), but rather by stochastic dynamical interactions (i.e. scattering) with other protoplanets in the disc, as well as with the binary itself (G. A. L. Coleman 2024; M. Teasdale & D. Stamatellos 2024; A. Čalović et al. 2026; S. Nayakshin et al. 2026).

### 3.3 The statistical properties of protoplanets

#### 3.3.1 Mass distribution

Fig. 12 shows the distribution of initial and final masses of protoplanets. All models form protoplanets with initial masses of a few $M_J$ that are consistent the opacity mass limit for gas fragmentation which is thought to be 1-5 $M_J$ (C. Low & D. Lynden-Bell 1976; M. J. Rees 1976; A. P. Whitworth & D. Stamatellos 2006; A. C. Boley et al. 2010; K. M. Kratter, R. A. Murray-Clay & A. N. Youdin 2010b) and previous studies (A. C. Boley 2009; D. Stamatellos & A. P. Whitworth 2009a; A. Mercer & D. Stamatellos 2020). The CS model has a peak in formation mass distribution at 2–3 $M_J$, with only a small number of protoplanets forming with larger masses (Fig. 12a). In contrast, both circumbinary models have a peak in distribution at slightly lower masses, 1–2 $M_J$. ~55 per cent of the protoplanets in the CBF model form at 1–2 $M_J$, dropping to ~40 per cent at 2–3 $M_J$. The CBR protoplanets, while following a similar trend, have 90 per cent of their total number forming at 1–2 $M_J$, with the remaining 10 per cent forming with a larger mass. This small difference between the peaks of the distributions for different models is expected as the minimum mass for fragmentation, set by the opacity limit, increases with increasing temperature (e.g. A. P. Whitworth & D. Stamatellos 2006). The CBR discs are cooler and therefore the initial protoplanet masses are lower than those in the CS model. This is also true for the CBF model, despite the fact the discs have the same temperature profiles as those in the CS model. However, due to the presence of the binary, fragmentation happens at slightly larger disc radii, where the disc is cooler (see Section 3.3.2).

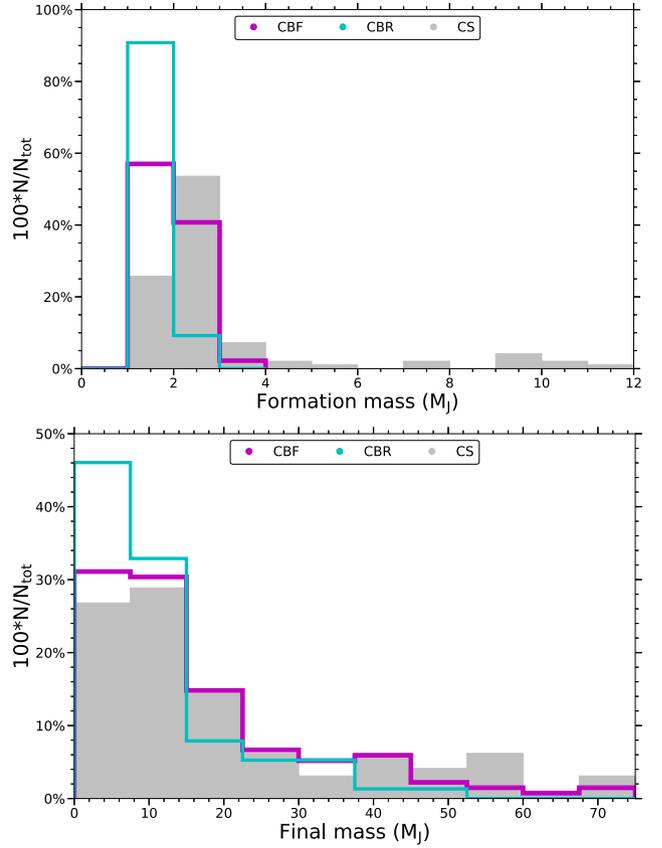

**Figure 12.** The mass distribution of protoplanets formed in the CS, CBF, and CBR model discs; the top panel (a) shows the distribution at the time of formation, and the bottom panel (b) at the end of the simulation.

In Fig. 12(b), which presents the final mass distribution, we see that in the CS model ~50 per cent of the protoplanets have masses below 13 $M_J$ (in the planetary-mass regime). The majority of CS model protoplanets have a final mass below ~16 $M_J$ (~60 per cent of all CS objects), with the remaining being within the BD mass-regime, i.e. no low-mass stars are formed in the disc. This is notably different to D. Stamatellos & A. P. Whitworth (2009a) who find ~70 per cent of the total objects being brown dwarfs and only ~3 per cent planets, with the remaining ~27 per cent having masses above the H-burning limit (80 $M_J$). This is because D. Stamatellos & A. P. Whitworth (2009a) use a significantly higher disc mass ($M_D = 0.7\,M_\odot$) and the objects that form in this disc are able to reach higher final masses than those present in our discs ($M_D \sim 0.2\,M_\odot$), as there is simply less mass to accrete. Moreover, we use the J. C. Lombardi et al. (2015) radiative approximation method that facilitates more efficient cooling that the D. Stamatellos et al. (2007b) method (see A. Mercer, D. Stamatellos & A. Dunhill 2018; A. K. Young et al. 2024).

Similarly to the CS model, ~54 per cent of CBF model protoplanets remain below 13 $M_J$ by the end of the simulations. The CBF model has a similar final mass distribution to the CS model, with the majority of protoplanets having a final mass below ~22 $M_J$, though a slightly larger proportion of protoplanets have a mass below 10 $M_J$.

The picture changes significantly in the CBR model, for which ~71 per cent of the protoplanets end up within the planetary-mass regime. The CBR model discs have the largest population of protoplanets below ~20 $M_J$, with 46 per cent having a final mass below 10 $M_J$. We therefore expect a larger fraction of planets to be able to form in realistic CB discs, compared to CS discs with similar properties.

### 3.3.2 Radial distribution

In the CS model the distribution of protoplanet's formation orbital radii peaks at ~75-100 au (~30 per cent of the protoplanets form in this range; Fig. 13a). In the CBF model, the distribution peaks at smaller radii, ~50-75 whereas in the CBR model most protoplanets (~70 per cent) forms within ~50-100 au. Very few protoplanets form within 50 au of the binary in the CBF and CBR models, whereas ~15 per cent form there in the CS model, demonstrating the significant impact that the binary has on disc fragmentation.

The distribution of the final orbital radii (Fig. 13b), shows that more protoplanets end up closer to the star in the CS model, whereas in the CB models they end up farther out, with the CBR model distribution peaking at around ~100 au; this also demonstrates the effect of the binary on the final orbital radii of the protoplanets. Most protoplanets end up within 200 au of their host star(s). This is to be expected as the initial disc radius is 120 au, and therefore it is unlikely that a large fraction of the protoplanets get scattered at a significant distance outside the disc. It is notable that a large percentage of protoplanets have a final radius beyond 100 au (around ~35 per cent for the CS model, 40 per cent for the CBF model and 44 per cent for the CBR model), reinforcing the idea that disc fragmentation favours the formation of wide orbit planets (e.g. A. C. Boley 2009). Protoplanets may end up even farther away than 200 au, due to planet–planet scattering as outward migration through interactions with the gas can only occur within the protostellar disc (see discussion in Section 3.2.3). Previous studies also find that reaching such a large final orbital radius is possible due to dynamical (planet–planet, or planet–binary) scattering (M. Teasdale & D. Stamatellos 2024; A. Čalović et al. 2026; S. Nayakshin et al. 2026).

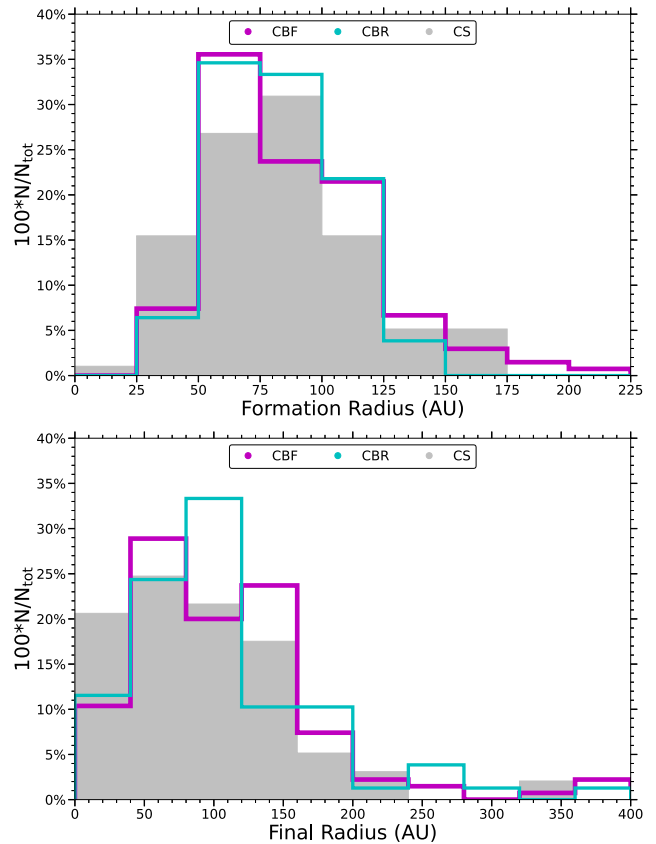

**Figure 13.** The radial distribution of protoplanets formed in the CS, CBF, and CBR model discs; the top panel (a) shows their radius at the time of formation, and the bottom panel (b) shows the radius at the end of the simulation.

Fig. 14 shows the distribution of the formation orbital radii of the protoplanets, separated between those with a final mass above and below 20 $M_J$. We adopt this division because objects with masses <20 $M_J$ are likely to evolve into planets, whereas those with masses >20 $M_J$ are more likely to become brown dwarfs. The distribution of CS protoplanets with mass above 20 $M_J$ peaks at 25–50 au (Fig. 14a), whereas for those with mass below 20 $M_J$ peaks at 75–100 au. The CBF and CBR models show very similar distributions for protoplanets with mass above 20 $M_J$ that they peak at larger radii (50–75 au) than the CS model distribution. The distributions of protoplanets with mass below 20 M$_J$ are very similar for all three models (Fig. 14b). Almost all of these form between 50 and 125 au, farther away from the central star(s) than the higher mass ones (cf. Figs 14a, b).

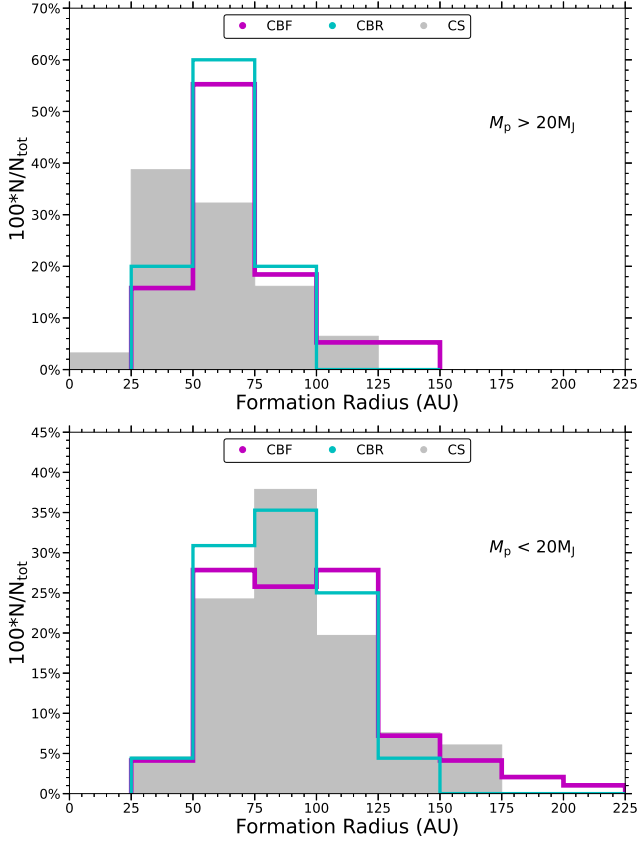

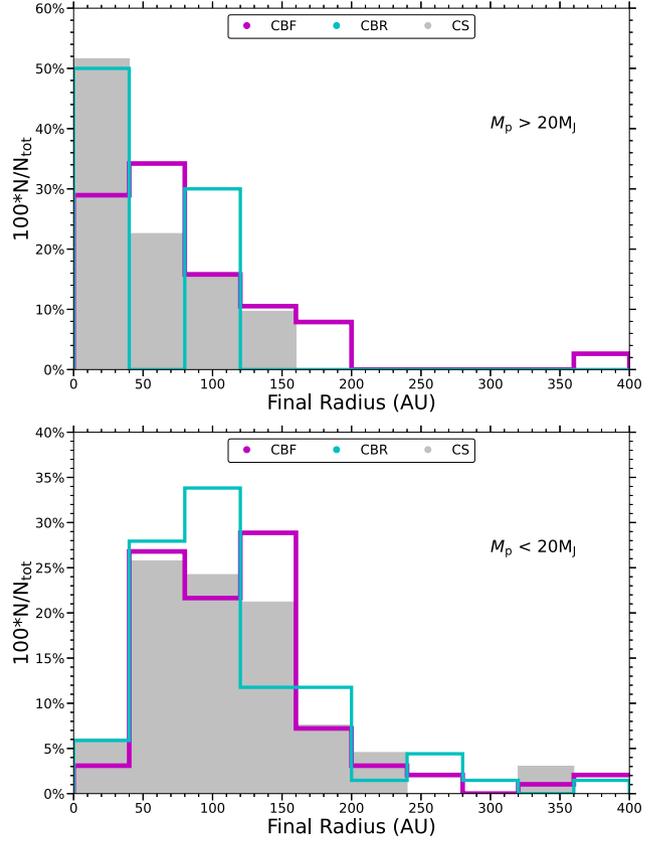

**Figure 14.** The initial orbital radii for fragments above and below 20 M$_J$ at the end of their respective simulations. The top panel (a) shows the formation radius above 20 M$_J$, and the bottom panel (b) shows the formation radius below 20 M$_J$.

**Figure 15.** The final orbital radii for fragments above and below 20 M$_J$ at the end of their respective simulations. The top panel (a) shows the final radius above 20 M$_J$, and the bottom panel (b) shows the final radius below 20 M$_J$.

The distributions of the final radii (Fig. 15) are quite similar for all three models with the only significant difference being that in CS discs more protoplanets with a final mass above 20 M$_J$ end up within ~45 au from the star. We find that in all three models, protoplanets with a final mass below 20 M$_J$ tend to end up at larger distances from their host star(s) than their more massive counterparts. This is because protoplanets that form closer to their star(s) tend to accrete more gas becoming massive, while at the same time scatter lower mass protoplanets away from the central region (see D. Stamatellos & A. P. Whitworth 2009a).

Comparing the corresponding peaks of the distributions in Figs 14 and 15 we see that higher mass protoplanets tend to migrate inwards, closer to the star, whereas lower mass protoplanets tend to migrate outwards (see also Section 3.2.3).

### 3.4 Ejection of protoplanets

The origin of free-floating planets (FFPs) remains an open question, with proposed formation channels including star-like formation via gravitational collapse (P. Padoan & Å. Nordlund 2002; P. Hennebelle & G. Chabrier 2008; A. Scholz et al. 2023; M. De Furio et al. 2025) or ejection from a protostellar disc (Y. Li et al. 2015,

2016; N. Miret-Roig 2023; G. A. L. Coleman & W. DeRocco 2025). FFPs exhibit a range of masses, from terrestrial (P. Mróz et al. 2020) to Jovian (T. Sumi et al. 2011), and are most readily detected in young clusters where they are yet to cool significantly enabling detections (N. Miret-Roig 2023). G. A. L. Coleman (2024) explored the formation of FFPs, finding that circumbinary systems are a good environment for ejections (see also A. Ćalović et al. 2026; S. Nayakshin et al. 2026). Furthermore, gravitational fragmentation has been found to produce FFPs and free-floating brown dwarfs, forming on wide-orbits before ejection (D. Stamatellos et al. 2011; A. Mercer & D. Stamatellos 2017). Here, we will explore whether fragmentation of CB discs is a more efficient mechanism in producing FFPs than fragmentation of CS discs.

A protoplanet is considered ejected from its host system when it fulfils three conditions: (i) its binding energy, E$_b$, is positive, (ii) its velocity is larger than the escape velocity, and (iii) its distance from the central star or the centre of mass of the binary is larger than 200 au. We calculate the binding energy of a protoplanet using

$$E_b = \frac{\mu\,v^2}{2} - \frac{G\,M_\star\,M_p}{r} - \frac{G\,M_p^F\,M_p}{r},\qquad(5)$$

where

$$\mu = \frac{M_\star\,M_p}{M_\star + M_p}.\qquad(6)$$

$v$ is the velocity of the protoplanet relative to the host star(s), $M_p$ the mass of the protoplanet, $M_\star$ the mass of the star (or total binary mass), and $r$ is the distance from the protoplanet to the star or, in the case of a binary, the distance to the centre of mass of the binary. The last term in the equation corresponds to the approximate gravitational potential of the disc at the end of the simulations; $M_D^F$ is the final mass of the disc (30 per cent of its initial mass). We ignore the contribution of protoplanets that are still bound to the stellar system. The escape velocity is calculated as

$$v_{esc} = \sqrt{\frac{2G(M_\star + M_D^F)}{r}}. \tag{7}$$

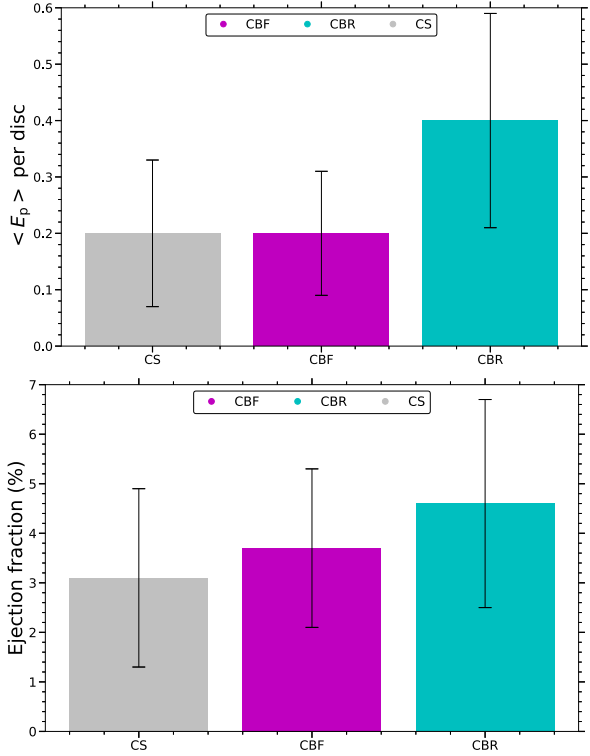

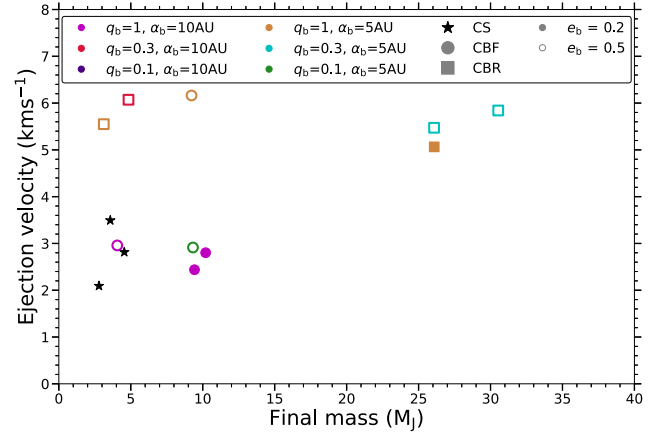

**Figure 17.** The final protoplanet mass plotted against their ejection velocity for the CS, CBF, and CBR models.

**Figure 16.** The top panel (a) shows the average number of ejections per disc, and the bottom panel (b) shows the ejection fraction, for each model. See Table 1 for the ejection fraction per model and disc.

The number of ejected protoplanets, $E_p$, is shown in Table 1. Of 341 protoplanets formed across the three disc models, 13 protoplanets are ejected, giving an combined ejection fraction of $4 \pm 1$ per cent. Fig. 16 presents the average number of ejections per disc and the ejection fraction for all three models. We see that the CBR model is likely more efficient in producing FFPs due to the larger number (on average) of protoplanets forming per disc in this model (see Fig. 7). D. Stamatellos & A. P. Whitworth (2009a) find a significantly higher ejection fraction of 15 per cent at the end of their SPH simulations; however, they have simulated more massive circumstellar discs that form a larger number of secondary objects, favouring dynamical interactions that lead to ejections.

Binary separation has a notable effect on protoplanet ejection fraction, with the closer binaries ($\alpha_b = 5$ au) having a larger ejection fraction of ~7 per cent, compared to the wider binaries ($\alpha_b = 10$ au), with ~2.5 per cent. This shows that tighter binary separations promote ejections due to stronger binary–planet encounters.

The initial binary mass ratio also affects the ejection fraction. We find that the ejection fraction increases with the binary mass ratio, going from ~1 per cent for $q_b = 0.1$, to ~4 per cent for $q_b = 0.3$, and to ~7 per cent for $q_b = 1$. This is in agreement with A. Ćalović et al. (2026), who find that a larger mass companion (higher binary mass ratio) leads to a higher ejection fraction.

Binaries with an eccentricity of $e_b = 0.2$ have an ejection fraction of ~3 per cent which is smaller than that for binaries with $e_b = 0.5$, which show an ejection fraction of ~5 per cent. Therefore, protoplanets that form around more eccentric binary systems are more likely to be ejected from their discs. The higher ejection fraction is due to a more prominent interaction between the binary and the protoplanets, which is in agreement with previous studies into binary–planet interactions (D. Veras & C. A. Tout 2012; M. Teasdale & D. Stamatellos 2024; A. Ćalović et al. 2026).

We find ejection velocities in the range 2–6.5 km s$^{-1}$ (see Fig. 17). Protoplanets ejected from CS discs have low ejection velocities, 2-4 km s$^{-1}$, which is consistent with D. Stamatellos & A. P. Whitworth (2009a). The presence of the binary results in typically higher ejection velocities. These velocities are consistent with A. Ćalović et al. (2026) and S. Nayakshin et al. (2026) who study ejections of disc-instability planets in binary systems, but not as high as those by G. A. L. Coleman (2024), who simulate ejections of core-accretion planets forming closer to the binary.

Our results demonstrate that increasing binary influence through larger separation, higher mass ratio, and greater eccentricity, leads to a higher probability of protoplanet ejection.

### 3.5 Comparison to observed exoplanets

Observations show a significant number of circumbinary planets at separations <10 au and at 300–1000 au (see Fig. 1). Comparing Figs 1 and 9(a), we find that our results are broadly consistent with observations of high-mass circumbinary planets (>1 M$_J$, i.e. the ones that can form though gravitational instability). The protoplanets formed in the realistic CB simulations have high-mass (see Fig. 9) and final orbital radii that fall within the range inferred from observations.

However, our models do not form close-in planets because the modelled binary separations (5 and 10 au) do not permit this; such planets are expected to form around much tighter binaries (see M. Teasdale & D. Stamatellos 2026, for the properties of binaries hosting circumbinary planets). Most of the protoplanets in the realistic CB model end up at separations around ∼100 au (see Fig. 15b), although a smaller fraction reach larger separations comparable to those observed. Further dynamical evolution of these systems is required to achieve better agreement with observations. The duration of our simulations (terminated when approximately 70 per cent of the protostellar disc has dissipated) does not capture the long-term evolution of the systems. As a result, there is limited time for protoplanet migration, either inwards or outwards, via disc-driven processes or through dynamical interactions with other protoplanets and the central binary.

## 4 CONCLUSIONS

We used the SPH code SEREN to study protoplanets formed through gravitational fragmentation in circumbinary and circumstellar discs. We performed three sets of simulations: the first covered circumstellar discs, the second covered circumbinary discs with the same temperature profile as the circumstellar discs (fiducial model), and the third set covered realistic circumbinary discs heated individually by each star of the binary (M. Teasdale & D. Stamatellos 2026). We varied the binary properties (separation, mass ratio, and eccentricity) to examine their effect on the formation and evolution of circumbinary protoplanets.

We find that discs around binaries with higher mass ratios and wider separations fragment faster than discs around binaries with lower mass ratios and smaller separations and faster than circumstellar discs. This is due to the effect of the gravitational influence of binary, which reaches farther out in the disc for wider and more eccentric systems. The asymmetric gravitational field of the binary provides a strong seed for the gravitational instability to grow faster. A larger number of protoplanets form in realistic circumbinary discs ($9 \pm 0.9$ protoplanets per disc) than in circumstellar discs ($7.5 \pm 0.8$ protoplanets per disc, as the disc is cooler and therefore easier to fragment. More protoplanets form in discs around wider binaries than in closer binaries. All these highlight the effect of the binary asymmetric gravitational field in promoting the growth of gravitational instability. At the end of the simulations, the protoplanets that remain bound to the stellar system and are closer to the host star(s) are more likely to have higher final masses than those at larger final orbital radii, due to a combination of higher mass growth in the gas-rich, inner disc region, and outward dynamical scattering of lower mass protoplanets.

The initial masses of the protoplanets that form in realistic circumbinary discs are smaller than those in circumstellar discs, as these form from fragmentation of cooler gas. Moreover, a larger fraction of protoplanets end up with masses within the planetary-mass regime rather than in the brown dwarf-mass regime (as they start from a lower mass and there is less gas in the disc to accrete), indicating that fragmentation of circumbinary discs is more likely to form planets than brown dwarfs or low-mass stars (cf. D. Stamatellos & A. P. Whitworth 2009a).

The distribution of formation radii is similar between circumstellar and circumbinary protoplanets; however, the presence of a binary creates a forbidden, dynamically unstable, region for fragmentation, resulting in almost all protoplanets forming outside 50 au. This propagates to the final radius distribution, where the peak occurs at around 100 au, farther out than in circumstellar discs. Once the long term dynamical evolution is taken into account (e.g. Y. Li et al. 2015, 2016) planets may end up on extremely wide orbits (M. Teasdale & D. Stamatellos 2024). We also find that a fraction of protoplanets may be ejected from the circumbinary disc, becoming free-floating (N. Miret-Roig 2023), due to dynamical interactions with other protoplanets and with the binary. This process is more efficient in circumbinary discs than in circumstellar discs (A. Ćalović et al. 2026; S. Nayakshin et al. 2026).

Our work strongly suggest that fragmentation of circumbinary discs is more efficient at forming gas giant planets than fragmentation of circumstellar discs around single stars of the same mass. This is because circumbinary discs (i) fragment at lower disc masses, and (ii) form a larger number of protoplanets per disc, such that the available disc mass is distributed among more objects.

Circumbinary disc fragmentation has been observed (J. J. Tobin et al. 2016; N. K. Reynolds et al. 2021), supporting the physical plausibility of this formation pathway, and our results indicate that it can also provide an efficient mechanism for producing FFPs through dynamical ejections. We therefore conclude that fragmentation driven by gravitational instability represents a viable and potentially significant formation channel for circumbinary gas giant planets.

## ACKNOWLEDGEMENTS

We thank the anonymous referee for their constructive review that helped improving the paper. The simulations were performed using the University of Lancashire High Performance Computing (HPC) and High Throughput Computing (HTC) facilities. The authors acknowledge support from the COST Action CA22133:PLANETS. DS acknowledges support from STFC grant ST/Y002741/1. We thank David Hubber for the development of SEREN. Surface density plots were produced using SPLASH (D. J. Price 2007). This work was partially performed using resources provided by the Cambridge Service for Data Driven Discovery (CSD3) operated by the University of Cambridge Research Computing Service (www.csd3.cam.ac.uk), provided by Dell EMC and Intel using Tier-2 funding from the Engineering and Physical Sciences Research Council (capital grant EP/T022159/1), and DiRAC funding from the Science and Technology Facilities Council (www.dirac.ac.uk).

## DATA AVAILABILITY

The simulation data used for this paper can be provided by contacting the authors.

## REFERENCES

Artymowicz P., Lubow S. H., 1994, ApJ, 421, 651
Betti S. K. et al., 2022, ApJ, 935, L18
Bodenheimer P., Pollack J. B., 1986, Icarus, 67, 391
Boley A. C., 2009, ApJ, 695, L53
Boley A. C., Hayfield T., Mayer L., Durisen R. H., 2010, Icarus, 207, 509
Ćalović A., Nayakshin S., Casewell S., Miret-Roig N., 2026, MNRAS, 545, staf2097
Chen C., Armitage P. J., Nixon C. J., 2026, ApJ, 997, L31
Christiansen J. L. et al., 2025, The Planetary Science Journal, 6, 186
Close L. M. et al., 2025, ApJ, 990, L9
Coleman G. A. L., 2024, MNRAS, 530, 630
Coleman G. A. L., DeRocco W., 2025, MNRAS, 537, 2303
Coleman G. A. L., Nelson R. P., Triaud A. H. M. J., 2023, MNRAS, 522, 4352
Cuello N. et al., 2026, A&A, 705, L16

Currie T. et al., 2022, Nat. Astron., 6, 751
Currie T. et al., 2025, ApJ, 990, L42
De Furio M. et al., 2025, ApJ, 981, L34
Delorme P. et al., 2013, A&A, 553, L5
Deng H., Mayer L., Helled R., 2021, Nat. Astron., 5, 440
Deng H., Mayer L., Meru F., 2017, ApJ, 847, 43
Doyle L. R. et al., 2011, Science, 333, 1602
Drążkowska J. et al., 2023, in Inutsuka S., Aikawa Y., Muto T., Tomida K., Tamura M., eds, ASP Conf. Ser. Vol. 534, Protostars and Planets VII. Astron. Soc. Pac., San Francisco, p. 717
Duchêne G., Bouvier J., Moraux E., Bouy H., Konopacky Q., Ghez A. M., 2013, A&A, 555, A137
Eriksson S. C., Asensio Torres R., Janson M., Aoyama Y., Marleau G.-D., Bonnefoy M., Petrus S., 2020, A&A, 638, L6
Fenton A., Stamatellos D., 2024, A&A, 682, L6
Fukagawa M., Tamura M., Itoh Y., Kudo T., Imaeda Y., Oasa Y., Hayashi S. S., Hayashi M., 2006, ApJ, 636, L153
Gammie C. F., 2001, ApJ, 553, 174
Goldreich P., Ward W. R., 1973, ApJ, 183, 1051
Guilloteau S., Dutrey A., Simon M., 1999, A&A, 348, 570
Haffert S. Y., Bohn A. J., de Boer J., Snellen I. A. G., Brinchmann J., Girard J. H., Keller C. U., Bacon R., 2019, Nat. Astron., 3, 749
Harris R. J. et al., 2018, ApJ, 861, 91
Heath R. M., Nixon C. J., 2020, A&A, 641, A64
Hennebelle P., Chabrier G., 2008, ApJ, 684, 395
Holman M. J., Wiegert P. A., 1999, AJ, 117, 621
Hubber D. A., Batty C. P., McLeod A., Whitworth A. P., 2011, A&A, 529, A27
Johnson B. M., Gammie C. F., 2003, ApJ, 597, 131
Keppler M. et al., 2018, A&A, 617, A44
Kley W., Haghighipour N., 2015, A&A, 581, A20
Kley W., Thun D., Penzlin A. B. T., 2019, A&A, 627, A91
Kratter K. M., Matzner C. D., Krumholz M. R., Klein R. I., 2010a, ApJ, 708, 1585
Kratter K. M., Murray-Clay R. A., Youdin A. N., 2010b, ApJ, 710, 1375
Kratter K., Lodato G., 2016, ARA&A, 54, 271
Langford A., Weiss L. M., 2023, AJ, 165, 140
Li J. et al., 2025, ApJ, 990, L70
Li Y., Kouwenhoven M. B. N., Stamatellos D., Goodwin S. P., 2015, ApJ, 805, 116
Li Y., Kouwenhoven M. B. N., Stamatellos D., Goodwin S. P., 2016, ApJ, 831, 166
Lombardi J. C., McInally W. G., Faber J. A., 2015, MNRAS, 447, 25
Low C., Lynden-Bell D., 1976, MNRAS, 176, 367
Marleau G.-D., Aoyama Y., Hashimoto J., Zhou Y., 2024, ApJ, 964, 70
Marley M. S., Fortney J. J., Hubickyj O., Bodenheimer P., Lissauer J. J., 2007, ApJ, 655, 541
Martín E. L., Cabrera J., Martioli E., Solano E., Tata R., 2013, A&A, 555, A108
Marzari F., Scholl H., 2000, ApJ, 543, 328
Marzari F., 2019, Galaxies, 7, 84
Marzari F., Thébault P., Scholl H., 2008, ApJ, 681, 1599
Mercer A., Stamatellos D., 2017, MNRAS, 465, 2
Mercer A., Stamatellos D., 2020, A&A, 633, A116
Mercer A., Stamatellos D., Dunhill A., 2018, MNRAS, 478, 3478
Meru F., Bate M. R., 2012, MNRAS, 427, 2022
Miranda R., Muñoz D. J., Lai D., 2017, MNRAS, 466, 1170
Miret-Roig N., 2023, Ap&SS, 368, 17
Mizuno H., 1980, Prog. Theor. Phys., 64, 544
Moe M., Di Stefano R., 2017, ApJS, 230, 15
Mróz P. et al., 2020, ApJ, 903, L11

Muñoz D. J., Miranda R., Lai D., 2019, ApJ, 871, 84
Nayakshin S., 2017, Publ. Astron. Soc. Aust., 34, e002
Nayakshin S., Zhang L., Čalović, Lee H., Baruteau C., Meru F., Mayer L., 2026, MNRAS, 546, stag043
Offner S. S. R., Moe M., Kratter K. M., Sadavoy S. I., Jensen E. L. N., Tobin J. J., 2023, in Inutsuka S., Aikawa Y., Muto T., Tomida K., Tamura M., eds, ASP Conf. Ser. Vol. 534, Protostars and Planets VII. Astron. Soc. Pac., San Francisco, p. 275
Paardekooper S.-J., Leinhardt Z. M., 2010, MNRAS, 403, L64
Padoan P., Nordlund Å., 2002, ApJ, 576, 870
Penzlin A. B. T., Booth R. A., Nelson R. P., Schäfer C. M., Kley W., 2024, MNRAS, 532, 3166
Pierens A., Nelson R. P., 2008, A&A, 483, 633
Pollack J. B., Hubickyj O., Bodenheimer P., Lissauer J. J., Podolak M., Greenzweig Y., 1996, Icarus, 124, 62
Price D. J., 2007, Publ. Astron. Soc. Aust., 24, 159
Quarles B., Satyal S., Kostov V., Kaib N., Haghighipour N., 2018, ApJ, 856, 150
Radley I. C. et al., 2025, ApJ, 981, 187
Raghavan D. et al., 2010, ApJS, 190, 1
Rees M. J., 1976, MNRAS, 176, 483
Reynolds N. K. et al., 2021, ApJ, 907, L10
Rice W. K. M., Armitage P. J., Bonnell I. A., Bate M. R., Jeffers S. V., Vine S. G., 2003, MNRAS, 346, L36
Rice W. K. M., Lodato G., Armitage P. J., 2005, MNRAS, 364, L56
Ringqvist S. C., Viswanath G., Aoyama Y., Janson M., Marleau G.-D., Brandeker A., 2023, A&A, 669, L12
Sadavoy S. I. et al., 2024, A&A, 687, A308
Scholz A., Muzic K., Jayawardhana R., Almendros-Abad V., Wilson I., 2023, AJ, 165, 196
Stamatellos D., Hubber D. A., Whitworth A. P., 2007a, MNRAS, 382, L30
Stamatellos D., Inutsuka S.-i., 2018, MNRAS, 477, 3110
Stamatellos D., Maury A., Whitworth A., André P., 2011, MNRAS, 413, 1787
Stamatellos D., Whitworth A. P., 2009a, MNRAS, 392, 413
Stamatellos D., Whitworth A. P., 2009b, MNRAS, 400, 1563
Stamatellos D., Whitworth A. P., Bisbas T., Goodwin S., 2007b, A&A, 475, 37
Sumi T. et al., 2011, Nature, 473, 349
Teasdale M., Stamatellos D., 2023, MNRAS, 526, 6248
Teasdale M., Stamatellos D., 2024, MNRAS, 533, 2294
Teasdale M., Stamatellos D., 2026, MNRAS, 545, staf2129
Thun D., Kley W., 2018, A&A, 616, A47
Tobin J. J. et al., 2016, Nature, 538, 483
Toomre A., 1964, ApJ, 139, 1217
van Capelleveen R. F. et al., 2025, ApJ, 990, L8
Veras D., Tout C. A., 2012, MNRAS, 422, 1648
Verhoeff A. P. et al., 2011, A&A, 528, A91
Vorobyov E. I., 2013, A&A, 552, A129
Wagner K., Apai D., Kratter K. M., 2019, ApJ, 877, 46
Wagner K. et al., 2018, ApJ, 863, L8
Wang J. J. et al., 2020, AJ, 159, 263
Whitworth A. P., Stamatellos D., 2006, A&A, 458, 817
Winn J. N., Fabrycky D. C., 2015, ARA&A, 53, 409
Young A. K., Celeste M., Booth R. A., Rice K., Koval A., Carter E., Stamatellos D., 2024, MNRAS, 531, 1746
Zhou Y. et al., 2021, AJ, 161, 244
Zhou Y. et al., 2022, ApJ, 934, L13

This paper has been typeset from a TeX/LaTeX file prepared by the author.